\documentclass[aip,amsmath,amssymb,reprint,]{revtex4-1}

\usepackage{graphicx}
\usepackage{dcolumn}
\usepackage{bm}

\usepackage{cmap}
\usepackage{hyperref}
\usepackage[utf8]{inputenc}
\usepackage[T1]{fontenc}
\usepackage{mathptmx}

\usepackage[separate-uncertainty=true]{siunitx}

\usepackage{amsmath}

\usepackage{color}
\newcommand{\red}[1]{\textcolor{black}{#1}}
\newcommand{\blue}[1]{\textcolor{black}{#1}}
\usepackage{ulem}

\begin{document}

\title{Sub-Sm$^{-1}$ electromagnetic induction imaging with an unshielded atomic magnetometer}

\author{Cameron Deans}
\author{Luca Marmugi}
\email{l.marmugi@ucl.ac.uk}
\author{Ferruccio Renzoni}
\affiliation{Department of Physics and Astronomy, University College London, Gower Street, London, WC1E 6BT, UK}

\date{\today}

\begin{abstract}
Progress in electromagnetic induction imaging with atomic magnetometers has brought its domain to the edge of the regime useful for biomedical imaging. However, a demonstration of imaging below the required \SI{1}{\siemens\per\meter} level is still missing. In this Letter, we use an $^{87}$Rb radio-frequency atomic magnetometer operating near room temperature in an unshielded environment to image calibrated solutions mimicking the electric conductivity of live tissues. By combining the recently introduced near-resonant imaging technique with a dual radio-frequency coil excitation scheme, we image $\SI{5}{\milli\liter}$ of solutions down to \SI{0.9}{\siemens\per\meter}. We measure a signal-to-noise ratio of 2.7 at \SI{2}{\mega\hertz} for  \SI{0.9}{\siemens\per\meter}, increased up to 7.2 with offline averaging. Our work is an improvement of 50 times on previous imaging results, and demonstrates the sensitivity and stability in unshielded environments required for imaging biological tissues, in particular for the human heart.

\vskip 20pt
\begin{center}
This is a preprint version of the article appeared in Applied Physics Letters:\\
C. Deans, L. Marmugi, F. Renzoni, Appl. Phys. Lett. {\bf 116}, 133501 (2020) DOI: \href{https://doi.org/10.1063/5.0002146}{10.1063/5.0002146}.
\end{center}
\end{abstract}

\maketitle
Recent years have seen a vast increase in the applications of quantum \blue{technologies and, in particular, atomic magnetometers\cite{budker2007}} to the biomedical field. Notable examples include magnetocardiography \cite{belfi2007,bison2009,shah2013,jensen2018} and magnetoencephalography \cite{romalis2006,shah2013,boto2018}. Applications for monitoring the reactivity of the nervous system have been also reported \cite{iwata2019}. In all cases, the superior performance of the atomic magnetometers pushes existing technologies and diagnostic methods towards their fundamental limits.

However, mapping the electric conductivity of biological tissues \blue{ -- and in particular of the human heart --} is still an open issue. \red{To date, no diagnostic tool is capable of non-invasively mapping the conductivity} \blue{of cardiac tissue\cite{scirep2016}.} \red{Current investigations require the invasive \blue{recording} of activation potentials via surgically introduced electrodes. This does not allow direct mapping of conductivity, and presents issues due to the inconsistent adhesion of electrodes to the inner surface of the beating heart\cite{electrodes}.} 

Electromagnetic induction imaging -- often referred to as magnetic induction tomography\cite{griffiths2001} to highlight its tomographic capabilities -- has been proposed as a diagnostic tool for various conditions characterized by a variation or an anomaly in electric conduction \cite{griffiths1999, merwa2004, zolgharni2010, rabbit}. With this technique, eddy currents are excited in the specimen under investigation by an AC magnetic field (primary field). The response, containing information about the electric conductivity, electric permittivity, and magnetic permeability of the specimen, is detected via the magnetic field generated by the eddy currents (secondary field). One of the main limitations of this approach is the limited sensitivity of the magnetic field sensors in use. Therefore, until recently, detection and imaging were limited to relatively large samples\cite{griffiths2007residual}, in most cases well above useful volumes for medical applications. This issue was potentially solved by the demonstration of electromagnetic induction imaging with atomic magnetometers \cite{ol2014,apl2016, wickenbrock2016}. The higher sensitivity of the core sensor paved the path to applications with small volumes of low-conductivity materials\cite{marmugi2019,jensen2019}. Nevertheless, a direct demonstration of imaging of sub-Sm$^{-1}$ specimens in unshielded environments, as required for applications in the biomedical field\cite{marmugi2019}, is still lacking. 

In this Letter, we demonstrate electromagnetic induction imaging of sub-\SI{}{\siemens\per\meter} calibrated solutions with an $^{87}$Rb radio-frequency (RF) atomic magnetometer operating in an unshielded environment, near room temperature, without \red{the need of acquiring background or reference images}. The performance required for imaging samples mimicking the conductivity of the cardiac tissue\cite{heart1,heart2} ($\sigma_{\text{el}}\sim\SI{0.7}{\siemens\per\meter}$-$\SI{0.9}{\siemens\per\meter}$) is obtained by combining near-resonant imaging with dual RF coil driving. By using two RF coils in anti-phase, we enhance the sensitivity of the atomic magnetometer to changes in conductivity, without detrimental effects due to RF-induced power broadening. \red{This represents an alternative to augment the detected phase change by} \blue{increasing the magnetometer’s} \blue{ operational frequency: a larger primary field produces a corresponding increase of the amplitude secondary field produced by the object of interest\cite{griffiths2001,griffiths1999}.} To optimise the imaging performance, we minimize the measured phase noise with a redesigned approach to data collection and programmable near-resonant imaging\cite{marmugi2019}. In this way, we are able to obtain a proof-of-principle demonstration exactly matching the requirements for imaging of cardiac tissue\cite{scirep2016}.

\begin{figure}[htpb]
\includegraphics[width=\linewidth]{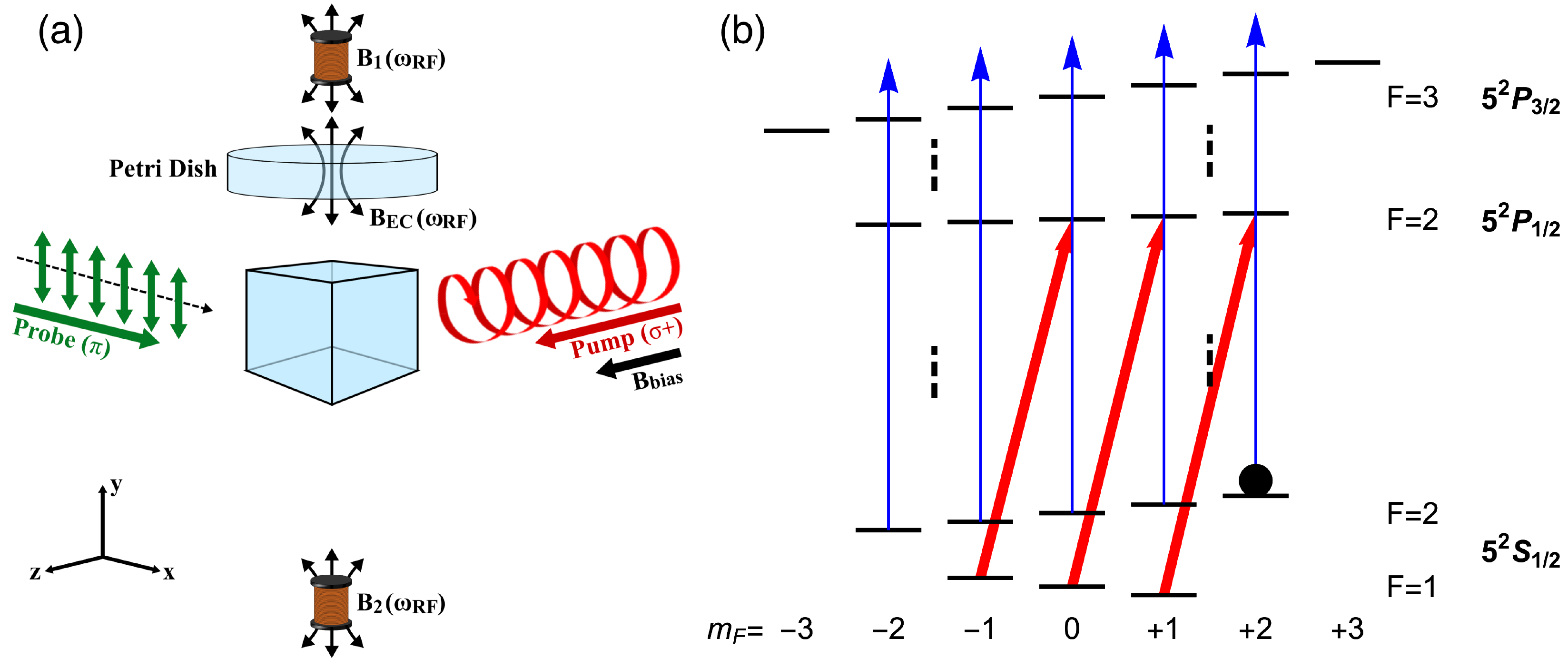}
\caption{\textbf{(a)} Schematic of the imaging setup with two RF coils. \textbf{(b)} Energy levels involved in the atomic magnetometer operation.}\label{fig:amemi}
\end{figure}

The experimental setup is a targeted evolution of a previous realization \cite{marmugi2019}. The atomic magnetometer is operated in an unshielded environment, by optically pumping $^{87}$Rb atoms in a \SI{25}{\milli\meter} quartz cell (buffer gas:  N$_{2}$, \SI{40}{\text{Torr}}) to the $|5^{2}$S$_{1/2}$F=2, m$_{\text F}$=+2$\rangle$ state via a $\sigma^{+}$-polarized laser beam at \SI{795}{\nano\meter} (\SI{+80}{\mega\hertz} detuning, \SI{330}{\micro\watt}) and a collinear DC bias field. This field (B$_{\text{bias}}$, in Fig.~\ref{fig:amemi}(a)) is locked to the desired value, using a three-axis fluxgate (Bartington MAG690), \SI{46}{\milli\meter} away from the center of the vapor cell\cite{rsi2018, witold2018} and a feedback loop. Passive compensation coils nullify DC stray magnetic fields and gradients. The readout of the atomic spin precession is obtained with a $\pi$-polarized probe beam, tuned \SI{+1300}{\mega\hertz} above the $|5^{2}$S$_{1/2}$F=2$\rangle \rightarrow |5^{2}$P$_{3/2}$F=3$\rangle$ transition Fig.~\ref{fig:amemi}(b). The probe's polarization plane is continuously monitored by a polarimeter, whose output is analyzed by a lock-in amplifier (Zurich Instruments HF2LI). Larmor precession of atomic spins is driven by an RF magnetic field ($\mathbf{B}_{1}(\omega_{\text{RF}})+\mathbf{B}_{2}(\omega_{\text{RF}})$ in Fig.~\ref{fig:amemi}(a)). The RF field is provided by a pair of \SI{100}{\micro\henry} coils, of diameter \SI{7.8}{\milli\meter}. They are placed at $\pm\SI{47}{\milli\meter}$ with respect to the center of the vapor cell, in an arrangement similar to those recently used in other atomic magnetometer setups\cite{bevington2019coils,jensen2019}, implementing the idea of Watson {\it et al.}\cite{watson2004}. The coils are antiparallel along the same axis. This creates two AC fields oscillating at the same frequency and with \blue{a} stable $\pi$ phase difference (anti-phase). The object to be imaged is placed between the top coil and the sensor, \SI{40}{\milli\meter} above its center (Fig.~\ref{fig:amemi}(a)).

The RF magnetic field -- $\mathbf{B}_{1}(\omega_{\text{RF}})+\mathbf{B}_{2}(\omega_{\text{RF}})$, the primary field -- induces eddy currents in the object. These produce \blue{the} secondary magnetic field ($\mathbf{B}_{\text{EC}}(\omega_{\text{RF}})$ in Fig.~\ref{fig:amemi}(a)), oscillating at the same frequency, and containing information about the properties of the object\cite{griffiths2001}. By moving the specimen, held in position by a motorized PLA and Nylon support, with respect to the sensor and measuring its response at desired position, 2D images are obtained.

\begin{figure}[htpb]
\includegraphics[width=\linewidth]{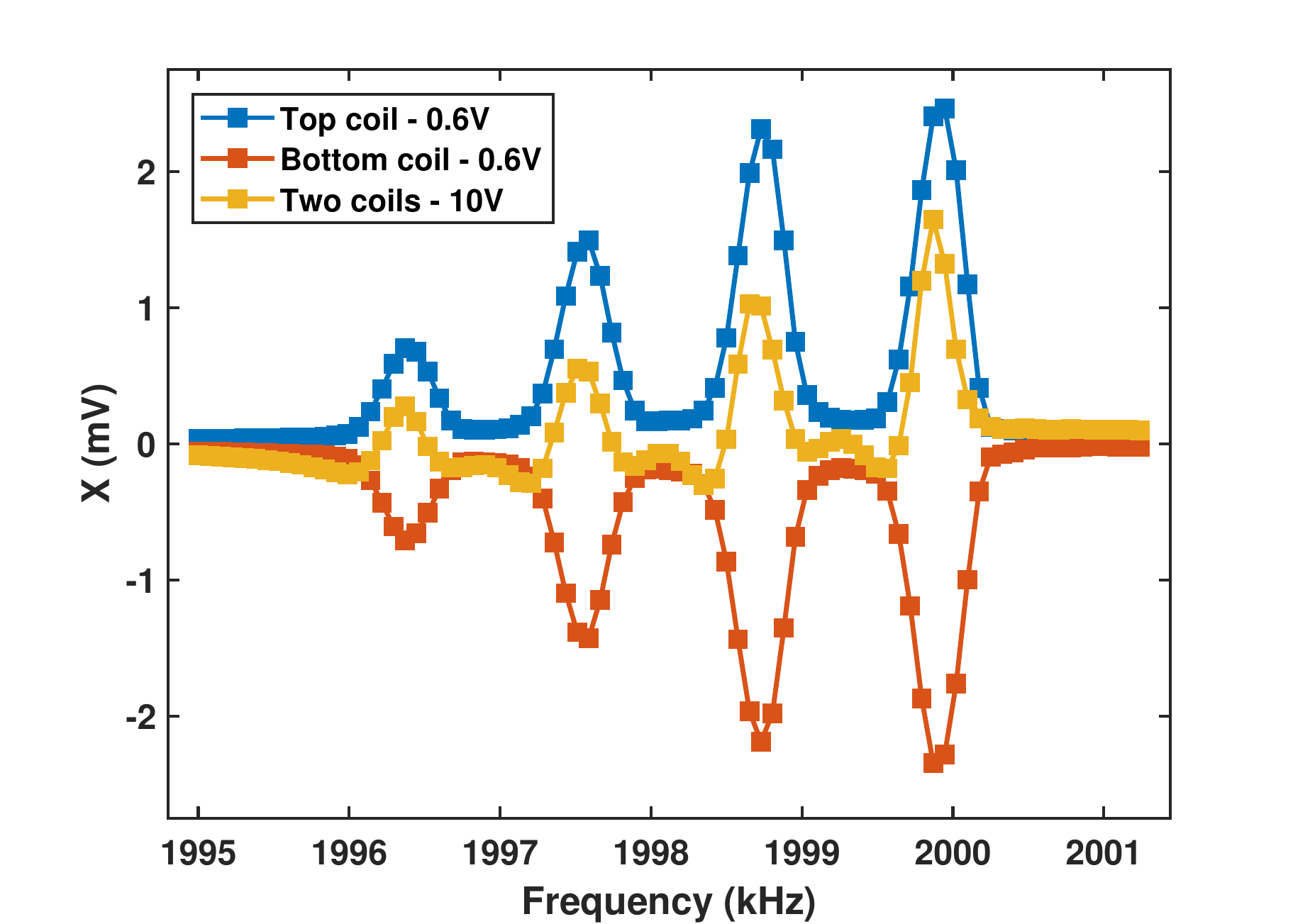}
\caption{Comparison of the in-phase (X) response with the top, bottom, and dual coils driving the AM for V$_{\text{RF}}$=\SI{0.6}{\volt}, \SI{0.6}{\volt}, and \SI{10}{\volt}, respectively.}\label{fig:resonance}
\end{figure}

One issue with electromagnetic induction imaging and atomic magnetometers are the contrasting requirements for the amplitude of the RF field. Increasing the amplitude of the primary field at the object's position proportionally increases the secondary field to be detected\cite{griffiths1999}. In contrast, increasing the amplitude of the driving field causes RF-induced power broadening, and a consequent reduction of the magnetometer's sensitivity\cite{rsi2018}. The dual RF coil, driven in parallel by the local oscillator of the lock-in amplifier, allows us to increase the amplitude of the RF field by more than 15 times at the object's level (the driving voltage for the RF coils is increased from V$_{\text{RF}}$=\SI{0.6}{\volt} to V$_{\text{RF}}$=\SI{10}{\volt}), without broadening the atomic resonances (Fig.~\ref{fig:resonance}). In this way, extremely low conductivity specimens can be imaged in unshielded environments, without the need to increase the frequency\cite{marmugi2019}. We did not completely cancel the response of the magnetometer (Fig.~\ref{fig:resonance}), instead creating a small imbalance in the currents flowing the two branches of the circuit (split ratio $\sim$51\%:49\%). \blue{By maintaining a residual RF field, we ensure the continuous and consistent operation of the magnetometer in unshielded environments, also in the absence of a field generated by the eddy currents in samples. This contributes to the stability of the measurements, whilst performing RF sweeps.} A different approach to long-term stability for electromagnetic induction imaging was recently demonstrated with a self-oscillating spin maser \cite{witold2019maser}.

Calibrated conductivity solutions, contained in Petri dishes (Star Lab CytoOne Dish, diameter \SI{35}{\milli\meter}, height \SI{10}{\milli\meter}), are used. The range of conductivities was chosen to mimic live biological tissues. \red{The size and the volume of the samples were chosen to represent a fraction of the size and volume of an adult human atrium \cite{atria}, where conductivity anomalies are expect to be found.} The specimens were independently assessed with a Jenway 4510 conductivity meter and a Jenway 027 213 Epoxy Bodied probe, at the temperature and time of imaging. We present results using three levels of conductivity: 1) \SI{9.1\pm0.1}{\siemens \per \meter} at \SI{20.1\pm0.1}{\celsius}, Reagecon CSKC100M (batch CS100M19H1); 2) \SI{4.5\pm0.1}{\siemens\per\meter} at \SI{20.0\pm0.1}{\celsius}, obtained with an NaCl (Sigma-Aldrich Redi-Dri reagent, $\geq99\%$ purity, 746398-500G) and de-ionized water (RS PRO De-ionized water, 251-3687) solution; 3) \SI{0.91\pm0.01}{\siemens \per \meter} at \SI{20.2\pm0.1}{\celsius}, Reagecon CSKC10M (batch CS10M19B2). Other tests were conducted with NaCl solutions at different concentrations. No noticeable difference in the response of different solutions with the same conductivity is detected. This suggests that the imaging process is independent of the chemical details of the solution and of the ionic carrier. The Petri dishes are filled with \SI{5.0\pm0.2}{\milli\liter} of solution, and then sealed to prevent evaporation.
\begin{figure}[htbp]
\includegraphics[width=\linewidth]{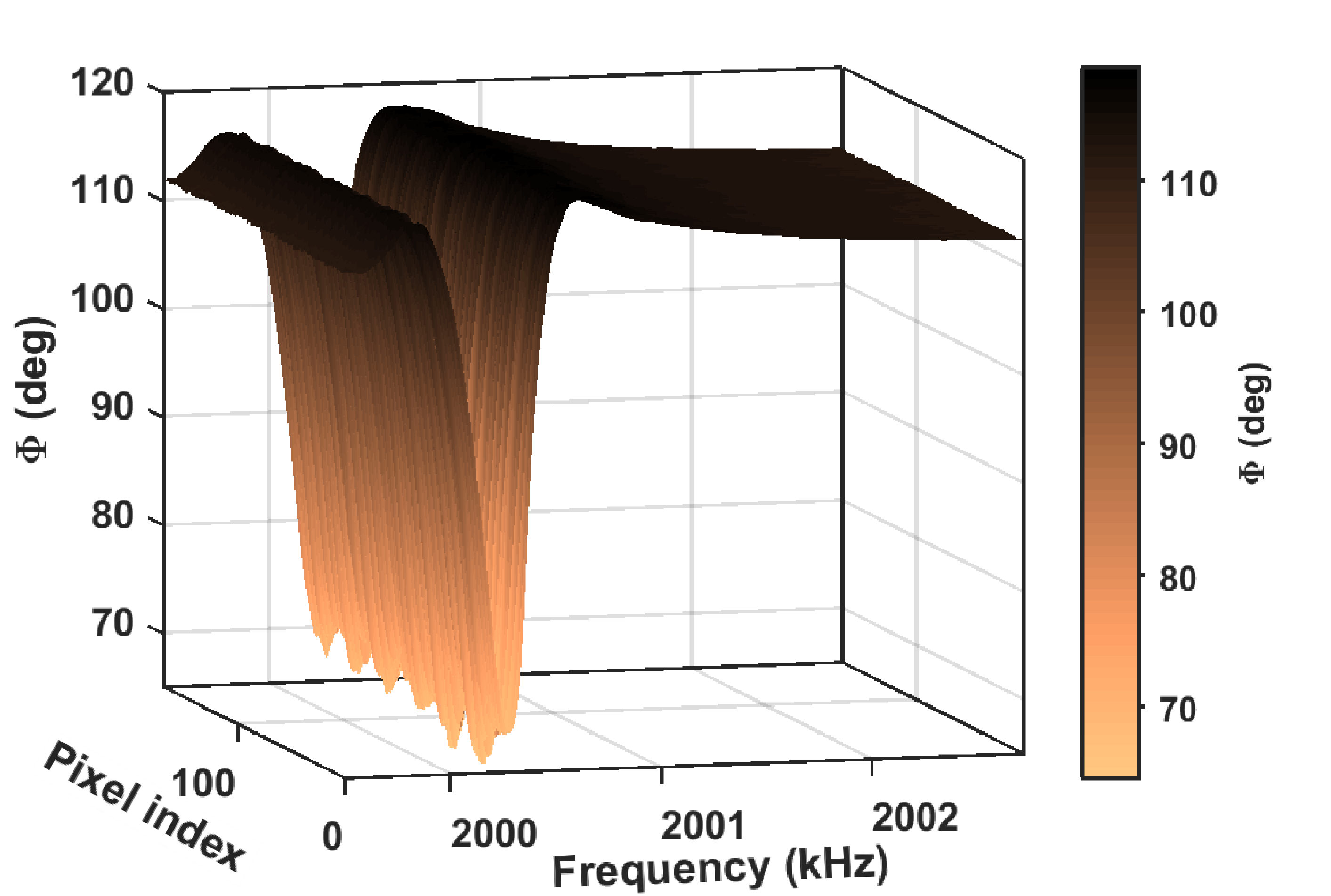}
\caption{\red{Phase ($\Phi$) spectra as a function of the RF frequency collected for a typical image. A complete image acquisition is constituted by X, Y, R, and $\Phi$ spectra per each pixel of the 2D imaging area.}}
\label{fig:signal}
\end{figure}
\begin{figure*}[htbp]
\includegraphics[width=0.9\linewidth]{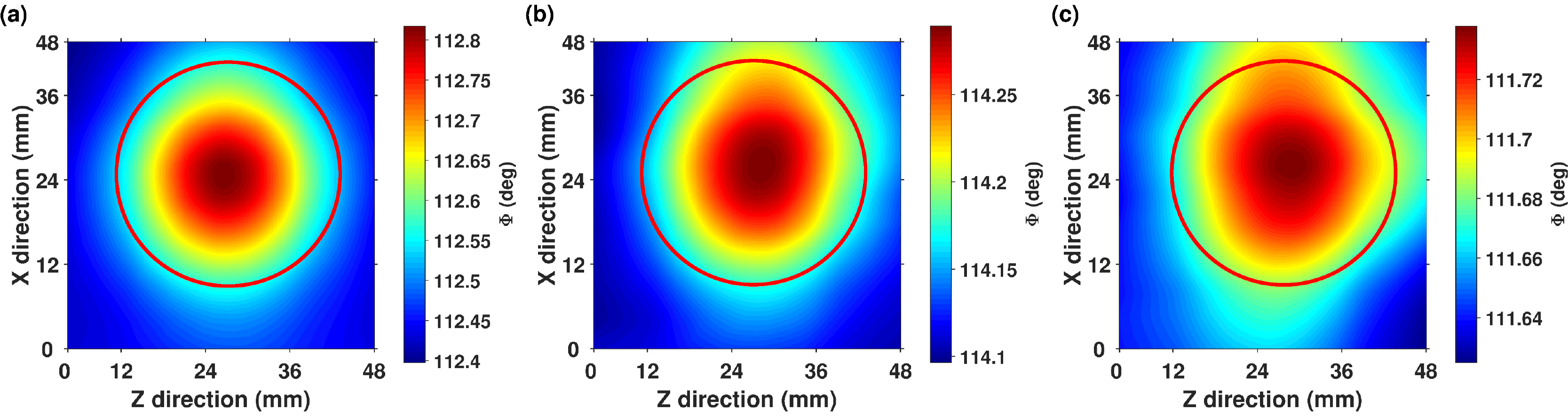}
\caption{Imaging of low-conductivity \SI{5}{\milli\litre} calibrated solutions. \textbf{(a)} Phase image of the \SI{9.1\pm0.1}{\siemens\per\meter} sample. \textbf{(b)} Phase image of the \SI{4.5\pm0.1}{\siemens\per\meter} sample. \textbf{(c)} Phase image of the \SI{0.91\pm0.01}{\siemens\per\meter} sample. \red{The red circles mark the position and the extension of the Petri dishes.}}
\label{fig:images}
\end{figure*}
\red{For imaging, we operate the magnetometer at around \SI{2}{\mega\hertz}.} \red{This frequency was chosen to allow direct comparison with previous results\cite{marmugi2019} and to highlight the advantages of the configuration presented in this Letter.} \red{To maximize the amount of the retained information, the system automatically acquires a full RF spectrum at each position of the image (i.e. for each pixel), by sweeping the RF field frequency across the atomic resonance, between \SI{1.995}{\mega\hertz} and \SI{2.001}{\mega\hertz}.} Five arrays, containing the swept frequency, the in-phase response (X), the quadrature response (Y), the amplitude of the response (R), and its phase lag ($\Phi$) are collected during the scan. For an $n\times m$ pixels scan and a $k$-points long frequency scan, five matrices are thus stored, each formed by $k$ columns and $(n\cdot m)$ rows. As an example, Fig.~\ref{fig:signal} illustrates the typical raw $\Phi$ data obtained by the system. In this case, each pixel is represented by a $\Phi$ spectrum of the atomic resonance recorded at that position. In this way, images are obtained for all parameters. \red{The small modulation appearing at the minima of Fig.~\ref{fig:signal} is attributed to residual magnetic inhomogeneities of the imaging volume, probably} \blue{caused by} \red{the motorized translational stage in use.} This approach allows the automatic implementation of near-resonant imaging\cite{marmugi2019} at a programmable detuning that minimizes the impact of magnetic noise in the parameter of interest. In the case of the $\Phi$ maps presented here, images can obtained at the near-resonant frequency exhibiting the minimum slope of the phase response. We found the optimum performance at \SI{2}{\mega\hertz} was +\SI{0.7}{\kilo\hertz} from the nominal resonance of the system. Multi-frequency imaging is also automatically attainable. Contrary to our previous demonstration\cite{marmugi2019}, here the optimization can be performed exclusively based on image quality without any intervention on the hardware. In addition, the complete set of raw data is fully retained for further analysis or performance enhancement.

\begin{figure}[htbp]
\includegraphics[width=\linewidth]{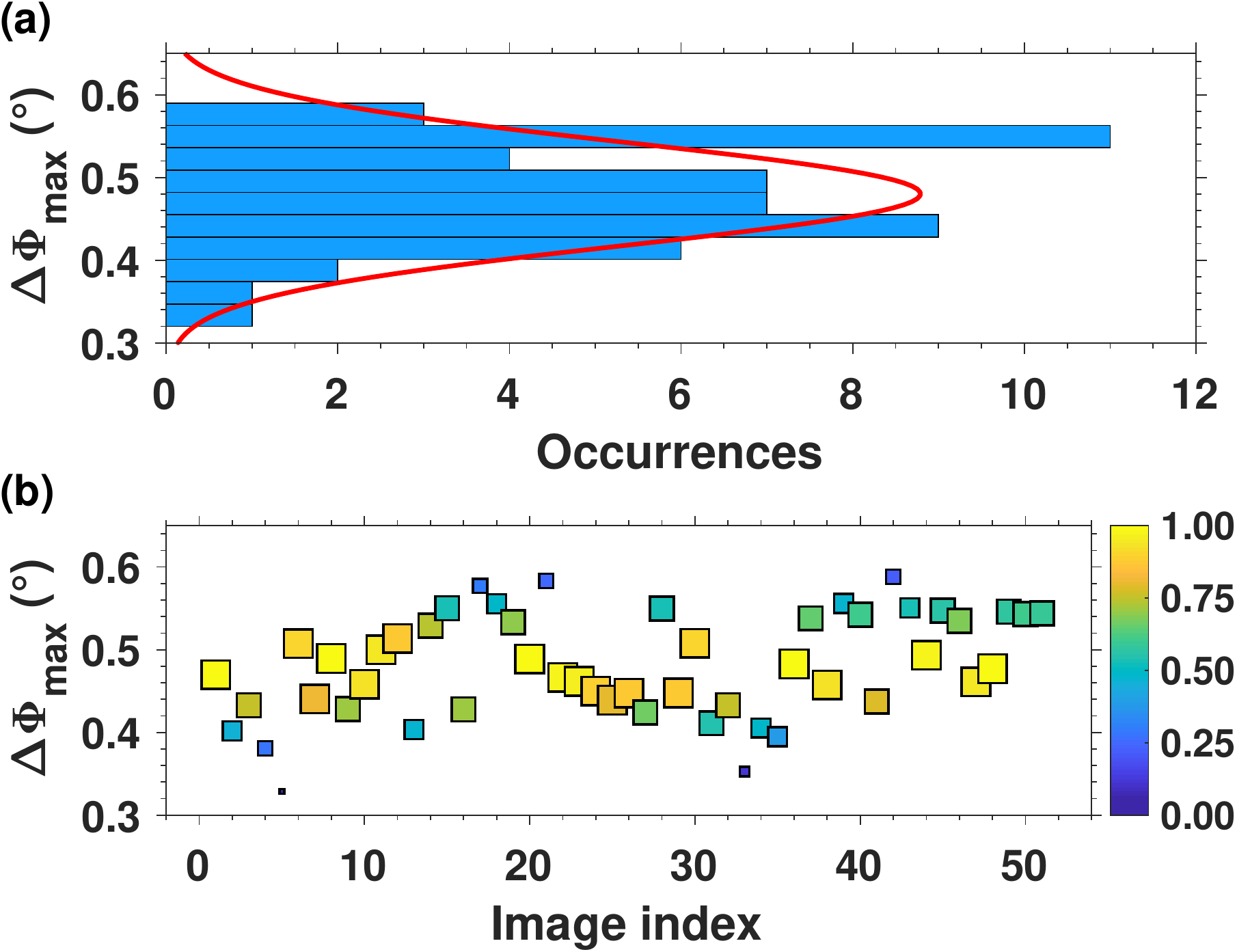}
\caption{\red{Stability of sub-\SI{}{\siemens\per\meter} imaging.} \textbf{(a)} Distribution of the maximum recorded phase change ($\Delta \Phi_{\text{max}}$) across 51 images images of the \SI{0.91}{\siemens\per\meter} sample. The red line is the fitting normal distribution. \textbf{(b)} Corresponding scatter plot. Size and color are proportional to the weight used for averaging. The colorbar indicates the weights $[0,1]$ attributed to the images.}
\label{fig:stats}
\end{figure}

In Fig.~\ref{fig:images}, we show the $\Phi$ maps of three Petri dishes containing \SI{5}{\milli\litre} of calibrated solutions, at \SI{9.1}{\siemens\per\meter} (a), \SI{4.5}{\siemens\per\meter} (b), and  \SI{0.91}{\siemens\per\meter} (c).  \red{Images in Fig.~\ref{fig:images} are obtained by weighted average of 43, 18, and 51 images (for (a), (b), and (c), respectively). Datafiles were collected consecutively, at random intervals, under the same environmental conditions and with the same settings of the imaging systems. For a given conductivity, the weights are obtained with a normal distribution fit to the set of images (see Fig.~\ref{fig:stats} for further details). The obtained image is then interpolated with a cubic piecewise algorithm. A nearest-neighbor Gaussian filter (radius 2 pixels) is also applied} \blue{to aid visualization}.

The specimens under investigation are correctly detected, imaged, and located in the scene, within the \SI{4}{\milli\meter} step-size of the translational stage in use.  Minor distortions appear in the \SI{0.91}{\siemens\per\meter} sample (Fig.~\ref{fig:images}(c)). \red{These are attributed to residual magnetic fluctuations of the background, which are not entirely eliminated by the active stabilization of the bias field and the near-resonant imaging.}

This result demonstrates electromagnetic induction imaging with an atomic magnetometer of small volumes of sub-\SI{}{\siemens\per\meter} specimens. The solutions mimic the ionic conductivity typical of live tissues, where electrolytes move in the intra- and intercellular spaces\cite{heart1}. Figure \ref{fig:images}(c) \blue{mimics} imaging of the heart with non-invasive electromagnetic induction imaging with atomic magnetometers: the electric conductivity of the healthy cardiac tissue in the \SI{}{\mega\hertz} range varies\cite{heart1,heart2} between \SI{0.7}{\siemens\per\meter} and \SI{0.9}{\siemens\per\meter}. This represents the background conductivity on which anomalous structures supporting atrial fibrillation, and speculated to be characterized by an increase in conductivity\cite{scirep2016,deansspie}, emerge. The demonstrated performance of our instrument is thus sufficient for their detection. \red{We recall that, given the current the lack of diagnostic tools dedicated to atrial fibrillation, quantitative information on its supporting structures are not available\cite{scirep2016}. For this reason, an imaging system capable of mapping the conductivity, even at low resolution, of the cardiac tissue would be of primary importance. In this view, \blue{the results of Fig.~\ref{fig:images}(c),} together with the unshielded and automated operation, allows one to conclude that the feasibility of electromagnetic induction imaging of biological tissues is now fully demonstrated. If required, enhancement of the spatial resolution could be obtained via machine learning localization and} \blue{classification algorithms \cite{prl2018}.}

The minimum imaged conductivity is improved by more than 50 times with respect to previous results\cite{marmugi2019}, and by a factor 4 with respect to the recently reported detection of low-conductivity solutions in a shielded environment \cite{jensen2019}. However, when considering the sensitivity of electromagnetic induction imaging instrumentation, the effective volume supporting eddy currents and thus contributing to the generation of the secondary field is also relevant. The larger the volume exhibiting a given conductivity, the bigger is the signal produced. By assuming the most favourable conditions -- namely when the skin depth is larger than the thickness of the object and the transverse size of the primary field at the specimen's surface is of the same order of the object's size -- one can simply take into consideration its physical volume ($V$). Following this reasoning, we introduce a figure-of-merit to characterize the absolute conductivity sensitivity of systems performing electromagnetic induction imaging:

\begin{equation}
\mathcal{S}\equiv \sigma_{\text{el}}\cdot V~\left[ \SI{}{\siemens\meter\squared} \right],\label{eqn:merit}
\end{equation}

\noindent where $\sigma_{\text{el}}$ is the electric conductivity of the object of interest. Based on Eq.~\ref{eqn:merit}, our record performance with the $\sigma_{\text{el}}$=\SI{0.91}{\siemens\per\meter} sample of $V=\SI{5.0}{\milli\liter}$ yields $\mathcal{S}=\SI{4.55e-6}{\siemens\meter\squared}$, an improvement of more than two orders of magnitude over our previous work\cite{marmugi2019} ($\mathcal{S}=\SI{3.13e-4}{\siemens\meter\squared}$), and of an order of magnitude with respect to recent detection reports\cite{jensen2019} ($\mathcal{S}\approx\SI{3.0e-5}{\siemens\meter\squared}$). The  $\sigma_{\text{el}}$=\SI{4.5}{\siemens\per\meter} sample corresponds to $\mathcal{S}=\SI{2.25e-5}{\siemens\meter\squared}$, $\sim$26$\%$ better than previous works. 

$\mathcal{S}$ also highlights the potential of the unshielded electromagnetic induction imaging with an RF atomic magnetometer with respect to other approaches. \red{Our goal here is to compare bare performance with the state-of-the-art of electromagnetic induction imaging.} Detection of conductivities as low as \SI{0.5}{\siemens\per\meter} has been reported with standard coils\cite{griffiths2007residual}. However, the comparatively large volumes (\SI{2.3e3}{\milli\liter}) of the samples gives $\mathcal{S}=\SI{1.13e-3}{\siemens\meter\squared}$, three orders of magnitude worse than the level demonstrated here. \red{In a different context, current nitrogen-vacancy (NV) center magnetometers performing electromagnetic induction imaging\cite{wickenbrock2019} conveyed $\mathcal{S}\approx\SI{1.8e-1}{\siemens\meter\squared}$.} \blue{However, we note that assessing the smallest detectable object was not the scope of that work\cite{wickenbrock2019}, of previous ones\cite{marmugi2019,jensen2019,griffiths2007residual}, or of this Letter.}

We have also investigated the performance with various images of identical samples acquired across several hours. The single-image signal-to-noise ratio (SNR) is well above unity in all cases. For the \SI{9.1}{\siemens\per\meter} sample, we measured a SNR=13.4. For the \SI{4.5}{\siemens\per\meter} sample, SNR=8.0. Finally, the \SI{0.91}{\siemens\per\meter} sample's images exhibit a single-image SNR=2.74.

We have also explored the direct averaging of the images, to improve the signal-to-noise ratio and readability. This would not be necessary when computer algorithms or machine learning are used\cite{prl2018}. \blue{In this way,} long-term performance as well as stability and repeatability of imaging -- all critical features for imaging relying on unshielded atomic magnetometers -- can be tested and quantified. Averaging several images improves such numbers, as expected: the averaged signal-to-noise ratio, $\langle$SNR$\rangle$, is 35.5, 16.2, and 7.23 from the three conductivities used, with an improvement of around 2.6 times.

As an illustration of the stability of the imaging for the \SI{0.91}{\siemens\per\meter} sample, we show the distribution of the maximum phase difference measured across an image $\Delta \Phi_{\text{max}}$ for 51 \blue{unfiltered and not interpolated repetitions} (Fig.~\ref{fig:stats}(a)):
\begin{equation}
\Delta \Phi_{\text{max}}=\left( \Phi_{\text{max}} -  \Phi_{\text{min}} \right)_{\text{image}}~.
\end{equation}
This allows one to immediately identify issues with the instrument or of the imaging process. The dataset is well-fitted by a normal distribution, with mean value $\langle\Delta \Phi_{\text{max}}\rangle=\SI{0.48}{\degree}$ with \SI{95}{\percent} confidence interval $\left[\SI{0.46}{\degree},  \SI{0.50}{\degree}\right]_{0.95}$ and standard deviation $\sigma=\SI{0.06}{\degree}$ ($\left[\SI{0.05}{\degree},  \SI{0.08}{\degree}\right]_{0.95}$). In Fig.~\ref{fig:stats}(b) the effect of the weighted average for Fig.~\ref{fig:images}(c) is illustrated. Once the distribution of the images dataset is identified, it is used for weighting the contribution of the single images to the average. \red{Although the details of the mechanisms leading to atrial fibrillation are yet to be confirmed \cite{scirep2016}, one can claim that the phase sensitivity and stability demonstrated in Fig.~\ref{fig:stats} are sufficient to detect small anomalies of conductivity in the heart. For example, in case of a factor 2 increase in conductivity in a \SI{5}{\milli\liter} volume, we expect to detect a phase change around \SI{0.2}{\degree} with the instrument presented here. This is safely above the detection threshold and intrinsic variability of our imaging system ($\left[\SI{0.05}{\degree},  \SI{0.08}{\degree}\right]_{0.95}$).}

In conclusion, we have presented electromagnetic induction images of calibrated solutions of conductivities down to the sub-\SI{}{\siemens\per\meter} with an unshielded RF atomic magnetometer. The simultaneous use of near-resonance imaging and of a dual RF coil excitation provided the necessary sensitivity and stability to allow us to image \SI{5}{\milli\liter} of \SI{0.91}{\siemens \per \meter} solution, in magnetically unshielded environments. Our atomic magnetometer-based imaging system operates with sufficient stability to guarantee consistent imaging performance across several days. Based on our results, the feasibility of practical use of electromagnetic induction imaging with atomic magnetometers in the biomedical field is finally fully demonstrated. \red{In particular, our instrument meets the requirements for direct, non-invasive imaging of the cardiac tissue – for example for the diagnosis of atrial fibrillation \cite{scirep2016} – which now appears technically feasible.}

\begin{acknowledgments}
This work was supported by the UK Quantum Technology Hub in Sensing and Metrology, Engineering and Physical Sciences Research Council (EPSRC) (EP/M013294/1). We thank Dr Helena Wong and Prof Andrea Sella (UCL Department of Chemistry) for the loan of the conductivity meter and guidance on its operation.
\end{acknowledgments}

\nocite{*}
%

\end{document}